\documentclass[sigconf,screen]{acmart}
\RequirePackage{fix-cm}
\usepackage[normalem]{ulem}
\usepackage{graphicx}
\usepackage{multirow}
\usepackage{lscape}
\usepackage{longtable}
\usepackage{supertabular}
\usepackage{enumitem}
\usepackage{algorithmic}
\usepackage{textcomp}
\usepackage{pgfplots}
\usepackage{balance}
\pgfplotsset{compat=newest}
\usepackage{caption}
\usepackage{pgfplotstable}
\usepackage{comment}
\usepackage{booktabs}
\usepackage{url}
\usepackage{balance}
\usepackage{color, colortbl}
\usepackage{tabularx}
\usepackage{adjustbox}
\usepackage{dirtytalk}
\usepackage{csquotes}
\usepackage{ulem}
\usepackage{soul}
\usepackage{subcaption}
\usepackage{tikz}

\usepackage{framed}
\usepackage[utf8]{inputenc}

\usepackage{threeparttable, tablefootnote}
\setlist{nolistsep}

\usepackage{soul}
\usepackage{multirow}
\usepackage{pifont}
\usepackage{xspace}
\usepackage{epigraph} 

  \definecolor{ABlue}{HTML}{127bca}
 \definecolor{LHScolor}{HTML}{555555}
\usepackage{framed}

\definecolor{gray(x11gray)}{rgb}{0.75, 0.75, 0.75}
 	
\def\mybar#1{
  {\color{gray}\rule{#1cm}{8pt}}}

\usepackage{pifont}

\usepackage{hyperref}
\definecolor{gray}{gray}{0.8}
\definecolor{cyan}{rgb}{0.88,1,1}

\usepackage{cite}
\usepackage{algorithmic}
\usepackage{graphicx}
\usepackage{textcomp}
\usepackage{booktabs}

\usepackage[most]{tcolorbox}

\newcommand\toppkg{Top-500}

\newcommand\midpkg{Middle-500}
\newcommand\lowpkg{Bottom-500}

\newcommand{\nb}[2]{
	{
		{\color{black}{
				\fbox{\bfseries\sffamily\scriptsize#1}
				{\sffamily$\triangleright~${\it\sffamily #2}$~\triangleleft$}
	}}}
}
\begin{document}
\begin{sloppy}

\acmPrice{}
\acmDOI{10.1145/3611643.3613086}
\acmYear{2023}
\copyrightyear{2023}
\acmSubmissionID{fse23ivr-p22-p}
\acmISBN{979-8-4007-0327-0/23/12}
\setcopyright{acmlicensed}
\acmConference[ESEC/FSE '23]{Proceedings of the 31st ACM Joint European Software Engineering Conference and Symposium on the Foundations of Software Engineering}{December 3--9, 2023}{San Francisco, CA, USA}
\acmBooktitle{Proceedings of the 31st ACM Joint European Software Engineering Conference and Symposium on the Foundations of Software Engineering (ESEC/FSE '23), December 3--9, 2023, San Francisco, CA, USA}
\received{2023-05-03}
\received[accepted]{2023-07-19}

\title{Lessons from the Long Tail: Analysing \\ Unsafe Dependency Updates across Software Ecosystems}

\author{Supatsara Wattanakriengkrai}
\affiliation{%
  \institution{Nara Institue of Science and Technology}
  \city{Nara}
  \country{Japan}}
\email{wattanakri.supatsara.ws3@is.naist.jp}

\author{Raula Gaikovina Kula}
\affiliation{%
  \institution{Nara Institue of Science and Technology}
  \city{Nara}
  \country{Japan}}
\email{raula-k@is.naist.jp}

\author{Christoph Treude}
\affiliation{%
  \institution{University of Melbourne}
  \city{Melbourne}
  \country{Australia}}
\email{christoph.treude@unimelb.edu.au}

\author{Kenichi Matsumoto}
\affiliation{%
  \institution{Nara Institue of Science and Technology}
  \city{Nara}
  \country{Japan}}
\email{matumoto@is.naist.jp}

\begin{abstract}
A risk in adopting third-party dependencies into an application is their potential to serve as a doorway for malicious code to be injected (most often unknowingly).
While many initiatives from both industry and research communities focus on the most critical dependencies (i.e., those most depended upon within the ecosystem), little is known about whether the rest of the ecosystem suffers the same fate. Our vision is to promote and establish safer practises throughout the ecosystem.
To motivate our vision, in this paper, we present preliminary data based on three representative samples from a population of 88,416 pull requests (PRs) and identify unsafe dependency updates (i.e., any pull request that risks being unsafe during runtime), which clearly shows that unsafe dependency updates are not limited to highly impactful libraries.
To draw attention to the long tail, we propose a research agenda comprising six key research questions that further explore how to safeguard against these unsafe activities. This includes developing best practises to address unsafe dependency updates not only in top-tier libraries but throughout the entire ecosystem.
\end{abstract}


\maketitle

\section{Introduction}
Widespread adoption and use of third-party library dependencies in applications is evident by the massive scale of software ecosystems (e.g., NPM for JavaScript, Maven for Java, PyPI for Python).
For example, the NPM ecosystem supports more than 2.42 million JavaScript libraries\footnote{We use the term libraries instead of packages for consistency.} as of 2022~\citep{Librarie95:online}.
Libraries in these ecosystems form complex dependency relationships in which they depend on each other for functionality.
A side effect of this heavy reliance on such ecosystems is the potential for vulnerabilities in libraries to spread.
One such example is the Log4Shell vulnerability, which potentially left a large number of systems, including Fortune 500 companies~\citep{YfryTchs40:online}, open to malicious attacks.
Furthermore, in 2021 there was a 650\% year-on-year growth in security attacks by exploiting Open Source Software supply chains~\citep{Sonatype12:online}.

In response to recent hijacking of widely used libraries, such as \texttt{ua-parser-js}~\citep{faisalma21:online}, \texttt{coa}~\citep{vegedcoa43:online}, and \texttt{rc}~\citep{dominict15:online}, GitHub has intensified its efforts to enhance the security of these highly depended-upon libraries.
Other industry efforts include mandating two-factor authentication (2FA) for high-impact libraries, such as those with more than one million weekly downloads or 500 dependents~\citep{Top100np77:online}.
Efforts such as the Alpha and Omega Project specifically aim to help the most critical open source projects improve their security postures~\citep{AlphaOme74:online}.
Combined with the OpenSSF (Open Source Security Foundation), they target and select projects based on the work of the OpenSSF Securing Critical Projects Working Group using various techniques, such as expert opinions, data analysis, and the OpenSSF Criticality Score to identify the most critical open source software.

In this paper, we argue that identification and remedy of supply chain attacks should not be limited to high-impact libraries, as insecure practises (e.g., unsafe dependency updates) might span across the entire ecosystem. 
We define unsafe dependency updates as any pull request (PR) that risks being unsafe
during runtime.
The term \textit{``unsafe''} originates from programming languages~\citep{UnsafeRu41:online}, where the compiler is explicitly instructed \textit{``trust the code being executed without question''}. Our vision entails the implementation of protective measures across all tiers of the ecosystem.
We collected 88,416 PRs from 1,500 curated libraries from three tiers of NPM libraries based on their dependence. Using features proposed in previous work, we flag and analyse unsafe dependency updates.

Our preliminary results suggest that developer practises that occur along the long tail of libraries are equally important as those in high-impact libraries.
This finding suggests that developer practises should be re-evaluated to reduce the prevalence of unsafe dependency updates and to understand the reasons behind developers' choices to employ these unsafe practises.

\section{Preliminary Analysis}
Our preliminary study covers 1,500 JavaScript npm libraries.
\subsection{Detecting unsafe dependency updates}

Existing defences against supply chain attacks are often deemed impractical in realistic settings due to their high computational costs and scaling issues.
These include carefully reviewing all updates to verify any activity (e.g., the execution of untrusted external sources) related to dependencies, thus hardening the package infrastructure (e.g., transport security, two-factor authentication), program analysis, and anomaly detection~\citep{Thecompl84:online, Nikitin:2017, Livshits:2009, GarrettICSE-NIER19}. 
Garrett et al.~\citep{GarrettICSE-NIER19} suggested a lightweight method to detect potential supply chain attacks.
The authors argue that their features are lightweight and do not require code compilation, and refer to them as unsafe:
\begin{enumerate}
    \item \textit{adding new scripts} - As highlighted in the documentation~\citep{Aboutnpm57:online}, these scripts allow new scripts to be added to the configuration files, thus running the risk of introducing code or scripts from unknown sources that are different from the source code.
    \item \textit{accessing http, http2, https} - Send and receive Hypertext Transfer Protocol (HTTP) requests. As mentioned by Garrett et al.~\citep{GarrettICSE-NIER19}, there is a risk that unknown information could be introduced into the application.
    \item \textit{executing fs()} - Create, read, and write to the file system~\citep{Filesyst74:online}. This poses the potential to erase or modify the local file system.
    \item \textit{executing net module} - Based on the documentation~\citep{NetNodej20:online}, this module allows functions to open, listen, or write to stream-based TCP or IPC sockets. Exploits may enable the introduction of information into the application.
    \item \textit{executing eval()} - This is a module that can execute JavaScript code represented as a string~\citep{evalJava75:online}. This practise is unsafe because it allows introduced sources to execute at runtime.
    \item \textit{executing require()} - The require command~\citep{JavaScri65:online} allows modules that exist in separate files to be run at runtime. This can be risky if malicious modules are imported and executed. 
\end{enumerate}

\begin{table}[]
\centering
\caption{Collected dataset of update related PRs (with changes to \texttt{package.json} and/or \texttt{.js files})}
\label{tab:prevalence1}
\scalebox{.8}{
\begin{tabular}{lrrrrrrrr}
\toprule
\multirow{2}{*}{\textbf{Tiers}} & \multicolumn{2}{c}{\textbf{\# dependents}} & \multicolumn{1}{c}{\multirow{2}{*}{\textbf{\# PRs}}} & \multirow{2}{*}{\textbf{\begin{tabular}[c]{r}\# update \\ related PRs\end{tabular}}} & \multicolumn{3}{c}{\textbf{update related PRs per lib}} \\ 
\cmidrule(lr){2-3} \cmidrule(lr){6-8}
 & \multicolumn{1}{c}{\textbf{Max}} & \multicolumn{1}{c}{\textbf{Min}} & \multicolumn{1}{c}{} &  & \multicolumn{1}{c}{\textbf{Mean}} & \multicolumn{1}{c}{\textbf{Median}} & \multicolumn{1}{c}{\textbf{SD}}\\ \midrule
\toppkg & 850,362 & 46,221 & 40,941 & 29,283 & 58.56 &  10.0 & 345.90  \\
\midpkg & 1,000 & 572 & 39,341 & 28,088 & 56.17 & 14.0 & 116.86 \\
\lowpkg & 1 & 1 & 8,134 & 6,458 & 12.91 & 2.0 & 71.7  \\
\multicolumn{3}{l}{\textbf{Total}} & \multicolumn{1}{l}{\textbf{88,416}} & \textbf{63,829} & \multicolumn{1}{l}{} & \multicolumn{1}{l}{} & \multicolumn{1}{l}{} \\ \bottomrule
\end{tabular}
}
\end{table}

Various solutions have been proposed to protect against unsafe dependency updates based on these metrics.
This includes lightweight permission~\citep{Ferreira:icse2021}, machine learning techniques~\citep{sejfia:icse2022}, and has also led to other empirical studies in the area~\citep{duan:ndss2022}.
\textbf{In our work}, we use the six features proposed by Garrett et al.~\citep{GarrettICSE-NIER19} to detect prevalence in three tiers of NPM libraries. 

\begin{figure}
\centering
\includegraphics[width=.8\linewidth]{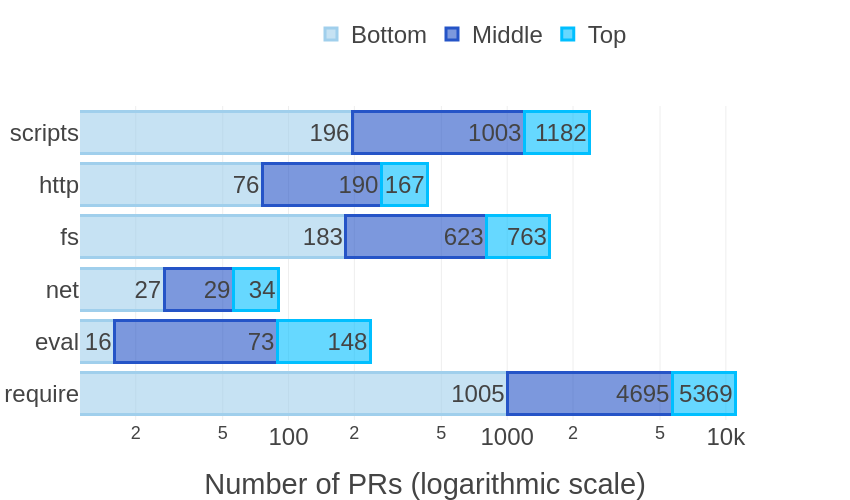}
\caption{Frequency of unsafe dependency updates}
\label{fig:rq1_keywords}
\end{figure}

\subsection{Library tiers}

Following prior work~\citep{GarrettICSE-NIER19}, we identify a PR as an \textit{update-related PR} if it involves changes to JavaScript files (\texttt{.js files}) and/or modifications to the \texttt{package.json} configuration file.
Note that the package.json file is used to record changes in dependencies.
We decided on a purposive sampling approach, taking an equal sample of libraries to represent the different populations within the ecosystem.
Our classifications are as follows:
\begin{itemize}
    \item \textbf{\toppkg} - The first tier of libraries are the most depended-upon libraries (outliers) within the NPM ecosystem. They are critical to the security supply chain. We ranked all the dependencies and selected the top 500 most depended-upon libraries. 
    \item \textbf{\midpkg} - The second tier of libraries is still highly depended-upon by the NPM ecosystem.
    According to GitHub, a high-impact library is defined as a library with more than one million weekly downloads or 500 dependents.
    To create this sample, we selected libraries that had a number of dependents between 500 and 1,000.
    \item \textbf{\lowpkg} - The final tier of libraries includes those with the least number of dependents. Libraries in this tier are the least depended upon, and thus we select libraries with at least one dependent.
\end{itemize}

Using previous work~\citep{Wattanakriengkrai:arxiv2022} as a starting point, we analysed 107,242 NPM libraries, which are then cross-referenced with the library dataset from libraries.io~\citep{Librarie95:online} used in previous studies~\citep{Golzadeh:2019, Zerouali:emse2022, Abdalkareem:emse2020}.
We collected 88,416 PRs from 1,500 sampled libraries based on their dependence as of March 2023.

Table \ref{tab:prevalence1} shows the details of the libraries selected for each tier. 
Filtering for update-related PRs (including package.json and/or .js files), we obtained 63,829 related PRs.
To flag unsafe dependency updates, we employ the six features identified earlier, using a simple regular expression to detect these features in lines added to PR code commits.
As shown in Figure 1, when applying the six features to detect unsafe dependency updates, we find that these updates are more likely to include the commands \texttt{require} and \texttt{new scripts}.
In contrast, the \texttt{net} and \texttt{eval} functions are less prevalent compared to the others.
All scripts and data can be found at 
\url{https://doi.org/10.5281/zenodo.6719258}.

\begin{table}[]
\centering
\caption{Unsafe dependency updates for each tier}
\label{tab:prevalence2}
\scalebox{.8}{
\begin{tabular}{lrrrrrr}
\toprule
\multirow{2}{*}{\textbf{Tiers}} &
\multirow{2}{*}{\textbf{\begin{tabular}[c]{r}\# unsafe \\ lib\end{tabular}}} &
\multirow{2}{*}{\textbf{\begin{tabular}[c]{r}\# unsafe \\ PRs\end{tabular}}}  &  \multicolumn{1}{c}{\textbf{Mean}} & \multicolumn{1}{c}{\textbf{Median}} & \multicolumn{1}{c}{\textbf{SD}} \\ \\ \midrule
\toppkg  & 405 (81\%) & 6,167 & 12.33 & 2.0 & 56.49 \\
\midpkg & 402 (80\%) & 5,212 & 10.42 & 3.0 & 27.87 \\
\lowpkg & 220 (44\%) & 1,086 & 2.17 & 0.0 & 12.58 \\
\multicolumn{2}{l}{\textbf{Total}}  \textbf{1,027} &\textbf{12,465} & \multicolumn{1}{l}{} & \multicolumn{1}{l}{} & \multicolumn{1}{l}{} \\ \bottomrule
\end{tabular}
}
\end{table}

\subsection{Prevalence}

Table \ref{tab:prevalence2} shows that we flagged a total of 12,465 PRs belonging to 1,027 libraries, which were distributed among the different tiers.
The evidence clearly shows that unsafe dependency updates are prevalent not only in the most impactful and dependent libraries, but across the other tiers within the ecosystem, including the \lowpkg~set of libraries.
For example, at least 81\% of the libraries in the \toppkg~category have at least one unsafe dependency update.
This is slightly higher than the \midpkg~(80\%).
Even \lowpkg~tier has almost 44\%, with at least one unsafe dependency update.

\begin{table}[!]
\centering
\caption{Unsafe dependency update outcomes per tier}
\label{tab:rq1_accept}
\scalebox{.8}{
\begin{tabular}{llrr}
\toprule
 &  & \multicolumn{2}{r}{\textbf{\# unsafe PRs}} \\ 
\toppkg & merged PRs & 4,333 & 70\% \mybar{0.7}\\
 & closed PRs & 1,508 & 24\% \mybar{0.24}\\
 & opened PRs & 326 & 6\% \mybar{0.06}\\
 & \textbf{Total} & \textbf{\begin{tabular}[c]{r}6,167\\ (21\% of update related PRs)\end{tabular}} & \\ \midrule
\midpkg & merged PRs & 3,704 & 71\% \mybar{0.71}\\
 & closed PRs & 1,242 & 24\% \mybar{0.24}\\
 & opened PRs & 266 & 5\% \mybar{0.05}\\
 & \textbf{Total} & \textbf{\begin{tabular}[c]{r}5,212 \\ (19\% of update related PRs)\end{tabular}} & \\ \midrule
\lowpkg & merged PRs & 863 & 79\% \mybar{0.79}\\ 
& closed PRs & 173 & 16\% \mybar{0.16}\\
 & opened PRs & 50 & 5\% \mybar{0.05}\\
 & \textbf{Total} & \textbf{\begin{tabular}[c]{r}1,086\\ (17\% of update related PRs)\end{tabular}} & \\ \bottomrule
\end{tabular}
}
\end{table}

\subsection{Acceptance} 

We calculate the acceptance rate of unsafe dependency updates by collecting the outcomes of the PRs after the review team has received them.
Typically, a PR is either accepted (i.e., merged) into the code base, closed without being merged into the code (i.e., closed) or still under review by the maintainers (i.e., opened).  
We compare the outcomes of the reviews of these unsafe dependency updates across the three tiers.

Table \ref{tab:rq1_accept} shows the acceptance of these unsafe dependency updates by the three tiers.
The results indicate that libraries in the least depended-upon tiers tend to accept and merge unsafe dependency updates.
As shown in the table, all tiers (i.e., \toppkg~at 70\%, \midpkg~at 71\%, and \lowpkg~at 79\%) have most of their unsafe dependency updates merged into the codebase. 

\subsection{Code Change Differences}

\begin{table}[]
\centering
\caption{The frequency of six unsafe dependency update types, comparing those labeled ``require attention''}
\label{tab:new_rq3}
\scalebox{.7}{
\begin{tabular}{@{}lrrrrr@{}}
\toprule
\textbf{} & \multicolumn{1}{l}{} & \multicolumn{2}{c}{\textbf{\begin{tabular}[c]{@{}c@{}}Unsafe PRs \\ with attention keywords\end{tabular}}} & \multicolumn{2}{c}{\textbf{\begin{tabular}[c]{@{}c@{}}Unsafe PRs \\ no attention keywords\end{tabular}}} \\
\textbf{Tiers} & \multicolumn{1}{c}{\textbf{\begin{tabular}[c]{@{}c@{}}\# merged \& \\ closed \\ PRs\end{tabular}}} & \textbf{\# PRs} & \textbf{PR types (\#)} & \textbf{\# PRs} & \textbf{PR types (\#)} \\ \midrule
Top-500 & 5,841 & 1,406 & \begin{tabular}[c]{@{}r@{}}\colorbox{orange}{Feature (1,120)}\\ \colorbox{orange}{Test Cases (838)}\\ \colorbox{orange}{Bug (725)}\\ Doc (546)\\ Refactoring (524)\\ Other (53)\end{tabular} & 4,435 & \begin{tabular}[c]{@{}r@{}}\colorbox{yellow}{Feature (2,797)}\\ \colorbox{yellow}{Bug (2,254)}\\ \colorbox{yellow}{Doc (1,832)}\\ Test Cases (1,310)\\ Refactoring (863)\\ Other (425)\end{tabular} \\ \midrule
Middle-500 & 4,946 & 1,136 & \begin{tabular}[c]{@{}r@{}}\colorbox{orange}{Feature (905)}\\ \colorbox{orange}{Bug (641)}\\ \colorbox{orange}{Test Cases (609)}\\ Doc (542)\\ Refactoring (252)\\ Other (48)\end{tabular} & 3,810 & \begin{tabular}[c]{@{}r@{}}\colorbox{yellow}{Feature (2,273)}\\ \colorbox{yellow}{Bug (1,525)}\\ \colorbox{yellow}{Doc (1,310)}\\ Test Cases (892)\\ Refactoring (482)\\ Other (438)\end{tabular} \\ \midrule
Bottom-500 & 1,036 & 149 & \begin{tabular}[c]{@{}r@{}}\colorbox{orange}{Feature (90)}\\ \colorbox{orange}{Bug (53)}\\ \colorbox{orange}{Test Cases (53)}\\ Doc (52)\\ Refactoring (42)\\ Other (20)\end{tabular} & 887 & \begin{tabular}[c]{@{}r@{}}\colorbox{yellow}{Feature (402)}\\ \colorbox{yellow}{Bug (283)}\\ \colorbox{yellow}{Doc (278)}\\ Test Cases (172)\\ Other (160)\\ Refactoring (109)\end{tabular} \\
\textbf{Total} & \textbf{11,823} & \textbf{2,691} & \multicolumn{1}{l}{} & \textbf{9,132} & \multicolumn{1}{l}{} \\\bottomrule
\end{tabular}
}
\end{table}

Since we employ the lightweight method from~\citep{GarrettICSE-NIER19}, it remains unconfirmed whether these unsafe dependency updates are indeed malicious.
Therefore, we investigate the differences in both code and textual content to better categorise these unsafe dependency updates.
We analyse both textual and committed file types to distinguish differences between the different tiers of unsafe dependency updates. 
We extract content from the titles and descriptions while identifying the types of file changes to determine the kind of changes that occurred.
Using a taxonomy of code changes from a previous study~\citep{Subramanian:2022}, two of the authors applied a saturation sampling approach to obtain the following coding. The codes consist of a combination of keywords and the types of files modified in the PR.
Noted that each keyword was removed non-textual symbols and punctuation, and applied lemmatization.
 \begin{itemize}
     \item Feature - \textit{keywords are: } integrate, add, feat, update, upgrade, support, dependency, feature, improve, version, automate, compatibility, bundle, improvement, bump
     \item Bug - \textit{keywords are: } avoid, fix, resolve, close, bug, solve, solution, issue, fixing
      \item Test Cases - \textit{keywords are: } test case, test, unit test, CI, continuous integration
      \item Refactoring - \textit{keywords are: } remove, unnecessary, refactor, performance, optimise
      \item Documentation - \textit{keywords are: } documentation, doc; \textit{files include: } .md files
      \item Other - \textit{files include: } .json only or anything that does not fall into the above categories
 \end{itemize}

To assess the impact of a PR, we follow related work~\citep{aom:emse2023} to identify keywords that attract developer attention (such as `attention', `breaking', and `performance').\footnote{The full list of the keywords is available at \url{https://doi.org/10.5281/zenodo.6719258}.}

Table \ref{tab:new_rq3} shows a consistent frequency count and content of unsafe dependency updates that attract attention and those that do not.
As shown, the number of unsafe dependency updates that require attention is relatively small across all tiers (i.e., 1,406 PRs for \toppkg, 1,136 PRs for \midpkg, and 149 PRs for \lowpkg), as these updates indicate critical problems for a library (e.g., performance issues and breaking changes).
Our results seem consistent across the tiers, with PRs classified as being new features and bugs being prevalent.
Interestingly, changes that required the maintainers attention included test cases, while those that did not (no attention), did include documentation changes.
We assume that test cases and documentation may indicate that the unsafe practises are justifiable, and intentional.

\section{Research Agenda}

Our research agenda is based on six key research questions that includes developing best practises to address unsafe dependency updates throughout the entire ecosystem.

\subsection{\textbf{Placing Safeguards}}
The first analysis findings indicate that unsafe dependency updates are prevalent across all tiers within the ecosystem. Thus, our aim is to establish a close connection between unsafe practises and their alternatives, based on existing research that has explored the exploitability of code~\citep{kang:issta2022} and how practitioners are using OSS libraries in their code~\citep{cispa3932}.

\textbf{Safer Alternatives:} Safer alternatives to these unsafe practises exist; but each has its own drawbacks that might not attract use. For example, an alternative to using the \texttt{require()} function is to instead use the \texttt{import()} function.
However, the key limitation is that the import function must be defined at the top of a class, while the require function can be executed at any point in the code~\citep{JavaScri86:online}. 
Similarly, for the \texttt{eval()} function, developers have documented the potential risks of using \texttt{eval()}~\citep{JavaScri60:online}; developers should consider using alternatives such as JSON.parse(), Function(), or templating engines.

\textbf{Adding Layers of Security:}
Another alternative is to add a layer of security by using appropriate HTTP security headers, which can help prevent some common attack vectors (for example, by using the helmet package).
Other solutions include implementing external source security certificate controls.
In terms of security access (i.e., unprotected access to files and directories on the Operating System), developers can design the program so that read/write access is restricted to certain files and directories only. Other options include sandboxing and the implementation of a permission system based on file and network access, as first studied by Ferreira et al.~\citep{Ferreira:icse2021}.

The first research question in our research agenda outlines the promises and perils of using alternative but safer update practises.

\begin{enumerate}
 
    \item What are the promises and pitfalls of using an alternative but safer implementation across the ecosystem?

\end{enumerate}

\subsection{Reasoning behind these Updates}
As mentioned in the previous section, there are safer alternatives and methods to remedy these unsafe practises; however, it remains unknown why developers persist in implementing these unsafe practises.
\begin{enumerate}[resume]
    \item What are the reasons why developers (open source and industry) are likely to trust (and accept) these unsafe dependency updates into their code?

\end{enumerate}

As the initial findings show that unsafe dependency updates are not limited to highly impactful libraries, it would be interesting to understand how the coding practises differ between the head and tail of the ecosystem.
For example, maybe a change of practices at the tail may propagate best practices at the head or vice versa.
As a community, the adoption and enforcement of practices may reap better results.
Researchers can further explore how the levels of unsafety may be different between updates in the head and
tail of the ecosystem.
This is essential because not all of the levels of insecurity/unsafety are equally concerning.
For example, some of the issues are just potential vulnerabilities that can hardly be exploitable in a realistic manner.
Therefore, our next research question aims to suggest learning from the long tail.

\begin{enumerate}[resume]

    \item How does the practise differ between the head and tail of dependencies, and how can we characterise those? 
\end{enumerate}

Answering these questions will provide a more comprehensive understanding of why these unsafe dependency updates exist.
Furthermore, we can complement and improve existing initiatives such as the criticality score~\citep{GitHubos50:online} and the OSS scorecard~\citep{GitHubos53:online}, which are currently being investigated by Zahan et al.~\citep{zahan:icse2023}.

\textbf{Automation for Safer Code:} The second analysis findings suggest that these unsafe dependency updates are being accepted.
Hence, developers are either unaware that these practises are unsafe or find them unavoidable due to time constraints and effort.
Actionable implications for researchers include tool support, such as code suggestion tools (e.g., automatic code completion), and refactoring existing unsafe code to make it safer in the appropriate situations.
Other tools include automatic detection of these unsafe dependency updates.
\begin{enumerate}[resume]
   
    \item How much effort is required to refactor unsafe code?

\end{enumerate}
\textbf{Utilising the Long Tail:}
The third analysis findings also show that unsafe dependency updates are feature-related and are accompanied by test cases or documentation, serving as a validation of trust that is consistent throughout the entire ecosystem.
Since this practise is prevalent, both researchers and security experts can target a larger collection of data and perform sampling across the long tail.

\begin{enumerate}[resume]
   
    \item What practises (open source and industry) of unsafe dependency updates can be trusted?
    \item What kind of validation is needed to gain trust from unknown sources that are using unsafe dependency updates?
 
\end{enumerate}

Answering these questions will provide us with actionable insights and also align with the OSSF goal of open source projects to adopt best practises when securing their code~\citep{GitHubos27:online}.

\section{Conclusion}
Studying the long tail of the software ecosystem demonstrates that unsafe practises exist in all tiers.
Therefore, the key to securing supply chains is not only to focus on highly impactful libraries but also to take into account existing practises and explore how the community as a whole can work together to create a safer and more resilient OSS supply chain.

\begin{acks}
Supported by JSPS KAKENHI Grant Numbers 20K19774, 23H03375 and 20H05706, and JST SICORP Grant Number JPMJSC2206.
\end{acks}

\balance
\bibliographystyle{ACM-Reference-Format}
\bibliography{filteredref}

\end{sloppy}
\end{document}